\documentclass[final,3p,times]{elsarticle}

\usepackage{amssymb}
\usepackage{epstopdf}
\usepackage{mathrsfs}
\usepackage{stmaryrd}
\biboptions{sort&compress}
\usepackage{graphicx,amsfonts,listings,xcolor,colortbl}
\usepackage[linesnumbered,ruled,lined]{algorithm2e}
\usepackage{multirow}
\usepackage{booktabs}
\usepackage{amsmath}
\usepackage{colortbl}
\usepackage{color}
\usepackage{rotating}
\usepackage{subfigure}
\usepackage[normalem]{ulem}

\begin{document}

\begin{frontmatter}



\title{Asymmetrically interacting dynamics with mutual confirmation from multi-source on multiplex networks}


\author[addr1]{Jiaxing Chen\fnref{equ}}
\author[addr1,addr3]{Ying Liu\corref{cor1}\fnref{equ}}
\ead{shinningliu@163.com}
\cortext[cor1]{Corresponding author}
\author[addr2,addr4]{Ming Tang\corref{cor1}}
\ead{tangminghan007@gmail.com}
\fntext[equ]{These authors contributed equally to this paper.}
\author[addr1]{Jing Yue}

\address[addr1]{School of Computer Science, Southwest Petroleum University, Chengdu 610500, P. R. China}
\address[addr2]{School of Physics and Electronic Science, East China Normal University, Shanghai 200241, China}
\address[addr3]{Big Data Research Center, University of Electronic Science and Technology of China, Chengdu 610054, P. R. China}
\address[addr4]{Shanghai Key Laboratory of Multidimensional Information Processing, East China Normal University, Shanghai 200241, China}

\begin{abstract}
In the early stage of epidemics, individuals' determination on adopting protective measures, which can reduce their risk of infection and suppress disease spreading, is likely to depend on multiple information sources and their mutual confirmation due to inadequate exact information. Here we introduce the inter-layer mutual confirmation mechanism into the information-disease interacting dynamics on multiplex networks. In our model, an individual increases the information transmission rate and willingness to adopt protective measures once he confirms the authenticity of news and severity of disease from neighbors status in multiple layers. By using the microscopic Markov chain approach, we analytically calculate the epidemic threshold and the awareness and infected density in the stationary state, which agree well with simulation results. We find that the increment of epidemic threshold when confirming the aware neighbors on communication layer is larger than that of the contact layer. On the contrary, the confirmation of neighbors' awareness and infection from the contact layer leads to a lower final infection density and a higher awareness density than that of the communication layer. The results imply that individuals' explicit exposure of their infection and awareness status to neighbors, especially those with real contacts, is helpful in suppressing epidemic spreading.
\end{abstract}

\begin{keyword}
multiplex network \sep epidemic spreading model\sep asymmetrically interacting dynamics \sep  mutual confirmation mechanism\sep microscopic Markov chain

\end{keyword}

\end{frontmatter}


\section{Introduction} \label{sec:intro}
The human beings have been fighting with epidemic disease for a long history. The recent coronavirus disease 2019 (COVID-19) has rapidly infected more than 25 millions of people and killed more than 800 thousands in just a few months~\cite{whosite}. Unraveling the mechanism that underlies the disease dynamics and adopting timely and accurate action are very important to save lives of many people and reduce the impact on the society and economics~\cite{zhang2020,kraemer2020,kissler2020,zhai2020}. In the longrun through social and economic activities, people form kinds of networks, such as the contact network, communication network, trade network, et al~\cite{newman2003}. The global spread of epidemics is thus a complex, network-driven process~\cite{anderson1992}. The underlying networks through which the disease transmits play a critical role in determining the spatiotemporal patterns of the dynamic process~\cite{hufnagel2004}. Using the complex network approach, some key properties of the epidemic process, such as the epidemic threshold, the disease arrival time, the critical transmission path, the spatial origin of spreading process, can be reliably predicted~\cite{claudio2010,dirk2013,romualdo2015}.

Nowadays with the rapid development of information technology, people communicate and obtain information much more easily and through multiple channels, e.g., kinds of online social networks, email, mobile phone and mass media. When epidemics outbreaks, the information about disease spreads and stimulates risk awareness among people. The preventive measures people adopt help to reduce infection and thus suppress the spreading of disease~\cite{funk2009}. The coevolving dynamics of information spreading and disease spreading can be modeled and analyzed as the asymmetrically interacting processes on the multiplex network, where the communication layer is formed by online social network and the physical contact layer is formed through persistent contacts in daily life~\cite{clara2013,granell2014,manlio2016}. The multiplex network is a particular kind of multilayer network where the same set of individuals form different layers. This multilayer approach has proved to be very successful in modeling the real complex systems where layers of networks are interrelated with each other~\cite{mikko2014,bocalleti2014,arruda2018}.

In the efforts to understand the information-disease coevolving spreading dynamics on multiplex network, a number of excellent models are proposed and some non-trivial phenomena are discovered which are substantially different from the independent spreading dynamics~\cite{clara2013,wang2014,guo2015,nico2017,moinet2018,yang2019}. For example, researchers demonstrated that there is a metacritical point for the oneset of the epidemics which is determined by the topology of the virtual network and the dynamics of information spreading~\cite{clara2013}. The outbreak threshold of epidemic is not affected by the information spreading, but an optimal information transmission rate may exist to markedly suppress the disease spreading~\cite{wang2016}. The time scale of information propagation doesn't affect the outbreak threshold, but an optimal relative timescale of information and disease spreading may reduce the epidemics incidence~\cite{wang2017,silva2019}. Some works define heterogenous disease or information transmission rate for nodes by taking the local or global infection or awareness density into consideration, where the awareness density is obtained from the communication layer and the infection density is obtained from the physical contact layer~\cite{zhang2014, pan2018,sagar2018}. In all the studies, the individuals receive information on one layer and transmit disease on the other layer.

In the real-world, people usually receive information from multiple source and the mutual confirmation of the information is likely to strengthen their willingness to accept the information and adopt responding activities. For example, before a customer makes a decision to purchase, he or she may get information from websites, social media, digital advertisements, catalogs, mobile and face-to-face recommendation. A consistent and positive evaluation of the focused product from multiple channels is probably to facilitate the customer's decision to purchase. In multilayer networks, the spreading processes on each layer are dynamically dependent on each other and their interactions are either interdependent or competitive~\cite{danziger2019}, which impact the activity of individual in different ways. In the information-disease coupling spreading dynamics, an individual can obtain the information on wether his neighboring nodes have been infected, either through the neighbors on the physical contact layer, e.g., from the neighbor's daily symptoms such as cough, fever, difficulty breathing, or through the neighbors on the communication layer, e. g., from the neighbor's blog contents or online shared pictures. These mutual confirmation on the infection of disease is likely to enhance the individual to accept and transmit the information about the epidemics. On the other hand, an individual can get to know wether his neighbors are aware of the disease either through the contact layer, from the neighbors' actions such as wearing mask, frequently cleaning hands or using hand sanitizers, intentionally maintaining at least one meter distance from others~\cite{whosite2}, or through the communication layer, e.g., the neighbor's attitude toward disease and safeguard in communications. These mutual confirmation of the information on disease may enhance the individual's willingness to adopt higher level of preventive measures, and thus further suppress the diffusion of disease.

In this manuscript, we introduce the mutual confirmation mechanism into the asymmetrically interacting spreading process on multiplex networks. By using the microscopic Markov chain approach (MMCA), we are able to capture the key quantities in epidemic spreading such as the epidemic threshold and the infection density in stationary state. The rest of the paper is organized as follows. In Section 2, the model with mutual confirmation mechanism is introduced. In Section 3, we use the Markov chain approach to analyze the epidemic threshold. In Section 4, we compare the MMCA predictions with the numerical simulation results and analyze the parameters. Finally in Section 5, we make conclusions.

\section{Model}
\subsection{The asymmetrically interacting spreading model on multiplex network}
\begin{figure}
\begin{center}
\includegraphics[width=13.5cm]{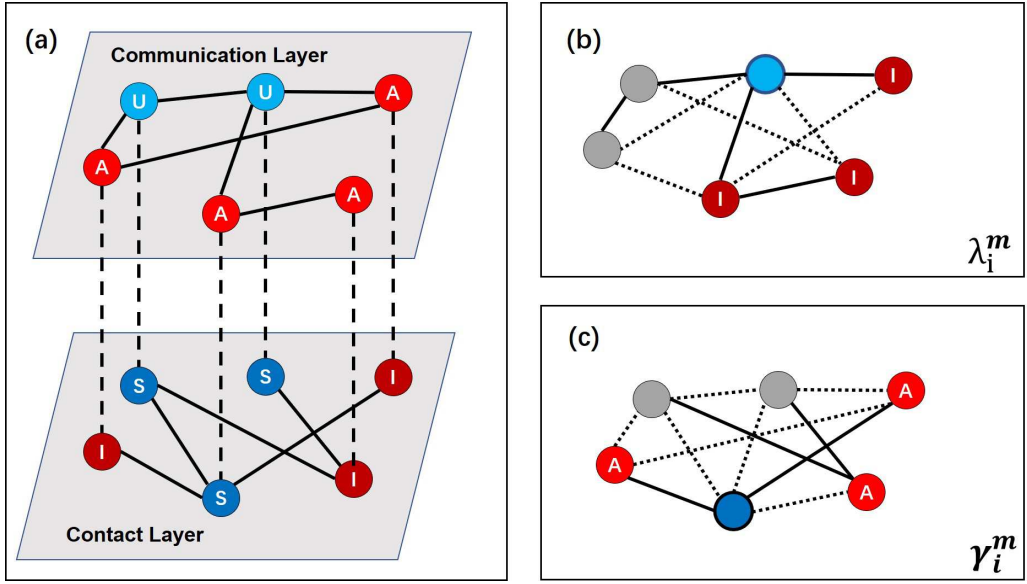}
\caption{The asymmetrically interacting spreading model with mutual confirmation. (a) The interacting spreading dynamics on a multiplex network. Layer A corresponds to the virtual communication layer where the information on disease spreads, which is a UAU dynamics. Layer B corresponds to the physical contact layer where disease transmits, which is an SIS dynamics. An infected node in layer B is automatically aware of the disease and its counterpart in layer A is in aware state. An aware node in layer A will increase its preventive measure against disease and its counterpart in layer B has a reduced rate of being infected as $\beta^A=\gamma\beta^U$. In the multiplex network, the nodes can be in one of the three states: US, AS and AI. Schematic diagram for calculating the information receiving rate $\lambda^{m}_i$ (b) and the attenuation factor $\gamma^{m}_i$ (c) under mutual confirmation mechanism. Solid and dotted lines represent connections in layer A and B respectively. For a U node noted light blue in (b), its proportion of infected neighbors is $2/3$ in layer A and and $1/2$ in layer B, so its $\omega_i^A=2/3$, $\omega_i^B=1/2$. For an S node noted dark blue in (c), its proportion of aware neighbors is $1$ in layer A and $1/3$ in layer B, so its $\nu_i^A=1$, $\nu_i^B=1/3$.}
\label{figure1}
\end{center}
\end{figure}

The information-disease interacting model is implemented on a multiplex network, as illustrated in Fig.~\ref{figure1}. The network consists of a virtual communication layer labeled as layer A and a physical contact layer labeled as layer B. The information spreading process is an unaware-aware-unaware (UAU) dynamics, where a node can be in either the unaware (U) state or aware state (A). The unaware individuals become aware once they have communicated with the aware individuals, and the aware individuals return to unaware state if they have forgotten or lost confidence on the news. The transmission rate of information is $\lambda$ and the recovery rate is $\delta$. In the physical contact layer, the classical susceptible-infected-susceptible (SIS) model is used, where an infected node propagates disease to each of its susceptible neighbor at disease transmission rate $\beta$, and infected nodes recover with rate $\mu$. An infected node in layer B is automatically aware of the disease and thus is in state A in layer A. An aware node is probably to adopt some preventive measures against disease, thus its infected probability is reduced. We use $\beta^A$ and $\beta^U$ to represent the disease transmission rate for nodes with and without awareness respectively, where $\beta^A=\gamma\beta^U$ and $\beta^U=\beta$. The attenuation factor $\gamma$ represents the extent of reduction of disease transmission rate. So the nodes in the multiplex network can be in one of the three states: unaware and susceptible (US), aware and susceptible (AS) and aware and infected (AI).

\subsection{The confirmation of multi-source information}
In the mutual confirmation mechanism, an individual can obtain the infection status of its neighbors, either from the communication layer or contact layer. A large proportion of infected neighbors from both layers mutually confirm that the disease is spreading and dangerous, which increases people's belief on the information and the information transmission rate is thus likely to be enhanced. Here we define the information receiving rate of node $i$ in the mutual confirmation mechanism as
\begin{equation}\label{lamda}
\lambda_i^{m}=\lambda(1+\theta_A\omega_i^A)(1+\theta_B\omega_i^B),
\end{equation}
where $\lambda$ is the basic information transmission rate, $\omega_i^A=\frac{\sum_ja_{ij}I_j}{k_i^A}$ is the proportion of infected neighbors in communication layer and  $\omega_i^B=\frac{\sum_jb_{ij}I_j}{k_i^B}$ is the proportion of infected neighbors in contact layer. $I_j=1$ if node $j$ is in infected states, otherwise $I_j=0$. The two parameters $\theta_A$ and $\theta_B$ quantify the confirmation strength of $\omega_i^A$ and $\omega_i^B$ in promoting the information transmission rate. Meanwhile, an individual is able to obtain the awareness status of his neighbors from both the communication layer and the contact layer. A large proportion of aware neighbors from both layers mutually confirm that the information about disease are widely accepted, which enhances the individuals' willingness to adopt higher level of preventive measures, therefore reduces the transmission rate of disease. Here we define the attenuation factor for disease transmission rate of node $i$ as
\begin{equation}\label{gamma}
\gamma_i^{m}=\gamma(1-\alpha_A\nu_i^A)(1-\alpha_B\nu_i^B),
\end{equation}
where $\gamma$ is the basic attenuation factor, $\nu_i^A=\frac{\sum_ja_{ij}A_j}{k_i^A}$ is the proportion of aware neighbors in communication layer and  $\nu_i^B=\frac{\sum_jb_{ij}A_j}{k_i^B}$ is the proportion of aware neighbors in contact layer. $A_j=1$ if node $j$ is in aware status, otherwise $A_j=0$. The two parameters $\alpha_A$ and $\alpha_B$ quantify the confirmation strength of $\nu_i^A$ and $\nu_i^B$ in reducing the disease transmission rate. When the four parameters $\theta_A=\theta_B=\alpha_A=\alpha_B=0$, then $\lambda_i^{m}=\lambda$, $\gamma_i^{m}=\gamma$, and the model reduces to the classical model without mutual confirmation mechanism. Symbols used in this paper are list in Table ~\ref{symbols}.
\begin{table}[htbp]
\begin{center}
\footnotesize
\caption{\label{symbols}Symbols used in the paper.}

\begin{tabular}{ll}
\hline
Symbol & Description \\
\hline
$A$($a_{ij}$)  &adjacent matrix for communication layer A (element in matrix $A$) \\
$B$($b_{ij}$)  &adjacent matrix for contact layer B (element in matrix $B$) \\
$k_i^A$($k_i^B$)  &degree of node $i$ in layer A(B) \\
$\beta$  &basic disease transmission rate \\
$\mu$  &disease recovery rate \\
$\lambda$  &basic information transmission rate\\
$\delta$  &information recovery rate\\
$\lambda^{m}_i$  &information receiving rate for node $i$ under mutual confirmation\\
$\beta^U$  &disease transmission rate for unaware node, $\beta^U=\beta$\\
$\beta^A$  &disease transmission rate for aware node\\
$\gamma$  &basic attenuation factor, $\beta^A=\gamma\beta^U$\\
$\gamma^{m}_i$  &attenuation factor for node $i$ under mutual confirmation\\
$\omega_i^A$($\omega_i^B$)  &proportion of infected neighbors in layer A(B) for node $i$\\
$\nu_i^A$($\nu_i^B$)  &proportion of aware neighbors in layer A(B) for node $i$\\
$\theta_A$($\theta_B$)  &confirmation strength of $\omega_i^A$($\omega_i^B$) in $\lambda^{m}_i$\\
$\alpha_A$($\alpha_B$)  &confirmation strength of $\nu_i^A$($\nu_i^B$) in $\gamma^{m}_i$\\
$p_j^A(t)$ & probability of node $j$ in A state at time $t$\\
$r_i(t)$ & probability of node $i$ not being informed by any neighbor at time $t$\\
$q_i^U(t)$ & probability of a U node not being infected by any neighbor at time $t$\\
$q_i^A(t)$ & probability of an A node not being infected by any neighbor at time $t$\\
$\rho^I$ & infection density in the stationary state\\
$\rho^A$ & awareness density in the stationary state\\
\hline
\end{tabular}
\end{center}
\end{table}
\section{Microscopic Markov chain approach}
\begin{figure}
\begin{center}
\includegraphics[width=13.5cm]{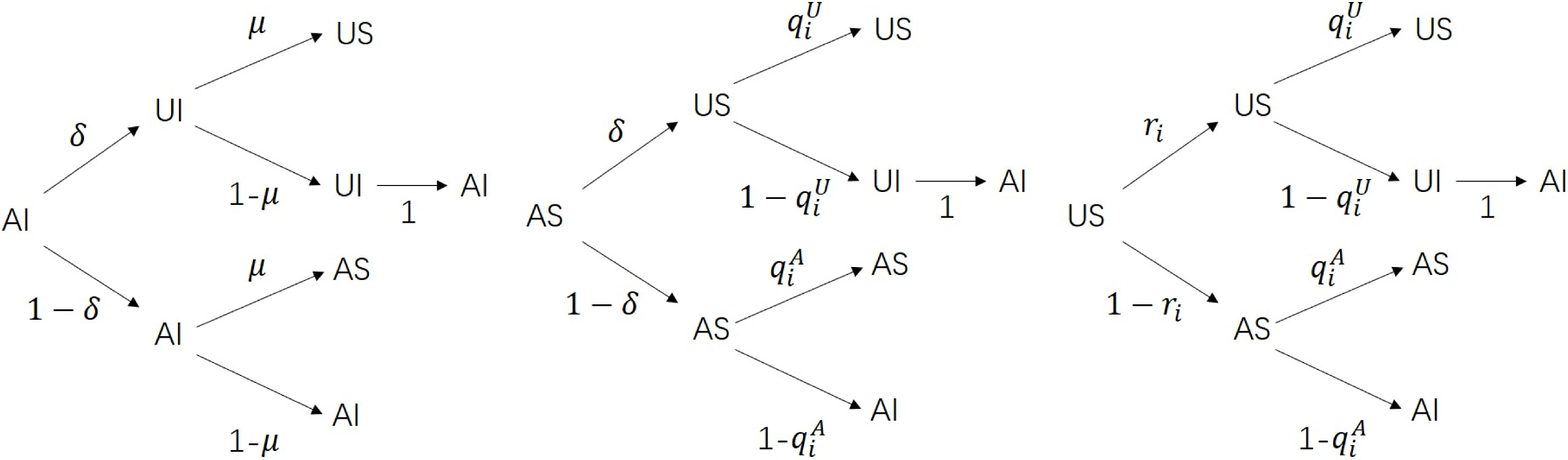}
\caption{Transition probability tree for the node states. There are three possible states for a node, US (unaware and susceptible), AS (aware and susceptible) and AI (aware and infected).}
\label{figure2}
\end{center}
\end{figure}

We use the microscopic Markov chain approach (MMCA)~\cite{clara2013} to describe the information-disease coupled spreading process in the proposed model. We denote the probability that node $i$ is in one of the three states US, AS and AI at time t as $p_i^{US}(t)$, $p_i^{AS}(t)$ and $p_i^{AI}(t)$ respectively. On the communication layer, the probability that a node is not informed by any neighbor at time t is
\begin{equation}\label{rt}
r_i(t)=\Pi[1-a_{ij}p_j^A(t)\lambda^{m}_i],
\end{equation}
where $\lambda^{m}_i=\lambda(1+\theta_A\omega_i^A)(1+\theta_B\omega_i^B)$ is the information receiving rate of node $i$ under mutual confirmation mechanism. On the contact layer, the probability of an unaware node or an aware node not being infected by any neighbor at time t are respectively
\begin{equation}\label{qu}
q_i^U(t)=\Pi[1-b_{ij}p_j^{AI}\beta^U]
\end{equation}
and
\begin{equation}\label{qa}
q_i^A(t)=\Pi[1-b_{ij}p_j^{AI}\beta^A],
\end{equation}
where $\beta^U$ and $\beta^A=\gamma^{m}_i\beta^U$ are the disease transmission rates for $U$ and $A$ nodes respectively, and $\gamma^{m}_i=\gamma(1-\alpha\nu_i^A)(1-\alpha_B\nu_i^B)$ is the attenuation factor for disease transmission rate under the mutual conformation mechanism.
The transition probability tree for node states is demonstrated in Fig.~\ref{figure2}.
Then the evolution equations of the three states in the coupled spreading dynamics can be written as
\begin{equation}\label{pUS}
p_i^{US}(t+1)\!=\!p_i^{AI}(t)\delta \mu\!+\!p_i^{US}r_i(t)q_i^U(t)\!+\!p_i^{AS}(t)\delta q_i^U(t),
\end{equation}

\begin{equation}\label{pAS}
p_i^{AS}(t+1)\!=\!p_i^{AI}(t)(1\!-\!\delta)\mu\!+\!p_i^{US}(t)[1\!-\!r_i(t)]q_i^A(t)\!+\!p_i^{AS}(t)(1\!-\!\delta) q_i^A(t)
\end{equation}
and
\begin{eqnarray}\label{pAI}
p_i^{AI}(t+1)&\!=\!p_i^{AI}(t)(1\!-\!\mu)\!+\!p_i^{US}\{[1\!-\!r_i(t)][1\!-\!q_i^A(t)]\!+\!r_i(t)[1\!-\!q_i^U(t)]\}\\\nonumber
&+p_i^{AS}(t)\{\delta[1\!-\!q_i^U(t)]\!+\!(1\!-\!\delta)[1\!-\!q_i^A(t)]\}.
\end{eqnarray}
To obtain the epidemic threshold, we use the stationary solutions of the system by letting $t\rightarrow \infty$ in Eqs. (\ref{pUS})-(\ref{pAI}), which satisfy \begin{equation}
p_i^{AI}(t+1)=p_i^{AI}(t)=p_i^{AI},
\end{equation}
\begin{equation}
p_i^{AS}(t+1)=p_i^{AS}(t)=p_i^{AS},
\end{equation}
\begin{equation}
p_i^{US}(t+1)=p_i^{US}(t)=p_i^{US}.
\end{equation}
Around the epidemic threshold, the number of infected nodes is negligible compared to the total population, thus the probability of nodes being infected can be assumed to be $p_i^{AI}=\epsilon_i\ll1$. Eqs. (\ref{qu}) and (\ref{qa}) can be approximated as
\begin{equation}\label{quapp}
q_i^{U}(t)=1-\beta^U\sum_jb_{ij}\epsilon_j,
\end{equation}
and
\begin{equation}\label{qaapp}
q_i^{A}(t)=1-\beta^A\sum_jb_{ij}\epsilon_j.
\end{equation}
Using the above conditions, we can simplify the stationary solution equations of Eqs. (\ref{pUS}-\ref{pAI}) as
\begin{equation}\label{pUSapp}
p_i^{US}=p_i^{US}r_i+p_i^{AS}\delta,
\end{equation}

\begin{equation}\label{pASapp}
p_i^{AS}=p_i^{US}[1-r_i]+p_i^{AS}(1-\delta),
\end{equation}

\begin{equation}\label{pAIapp}
\epsilon_i=\epsilon_i(1-\mu)+(p_i^{AS}\beta^A+p_i^{US}\beta^U)\sum_j\epsilon_jb_{ij}.
\end{equation}
By substituting $\beta^A=\gamma_i^{m}\beta^U$ into Eq. (\ref{pAIapp}), we obtain
\begin{equation}\label{uei}
\mu\epsilon_i=\beta^U(p_i^{AS}\gamma^{m}_i+p_i^{US})\sum_j\epsilon_jb_{ij}.
\end{equation}

Near the threshold $p^{AI}=\epsilon_i\ll1$, then
\begin{equation}\label{pa}
p_i^A=p_i^{AI}+p_i^{AS}\approx p_i^{AS}.
\end{equation}
In addition, there is $US$ node in the network, but no $UI$ node, so $p_i^{US}=p_i^U$. As $p_i^{AI}+p_i^{AS}+p_i^{US}=1$, then $p_i^U=1-p_i^A$. Eq. (\ref{uei}) can be rewritten as
\begin{equation}\label{ueiapp}
\mu\epsilon_i=\beta^U(\gamma_i^{'}p_i^A+p_i^U)\sum_j\epsilon_jb_{ij}.
\end{equation}
The Eq. (\ref{ueiapp}) can be written in the format of matrix as
\begin{equation}\label{ueimatrix}
\sum_j \{(p_i^U+\gamma_i^{'}p_i^A)b_{ij}-\frac{\mu}{\beta^U}\sigma_{ij} \} \epsilon_j=0,
\end{equation}
where $\sigma_{ij}$ is the element of identity matrix. When epidemic starts to pervade the network in stationary state, the epidemic threshold $\beta_c$ is the minimum value that satisfies Eq. (\ref{uei}). As a self-consistent equation, obtaining $\beta_c$ reduces to the eigenvalue problem~\cite{guo2015}. Let $\Lambda_{max}(H)$ be the largest eigenvalue of matrix $H$, where the elements of $H$ is
\begin{equation}\label{hij}
h_{ij}=(p_i^U+\gamma^{m}_i p_i^A)b_{ij}=\{[\gamma(1-\alpha_A \nu_i^A)(1-\alpha_B\nu_i^B)-1]p_i^A+1\}b_{ij}.
\end{equation}
Then the epidemic threshold can be written as
\begin{equation}\label{hij}
\beta_c=\frac{\mu}{\Lambda_{max}(H)}.
\end{equation}
In calculating $h_{ij}$, we iterate Eqs. (\ref{pUS}) and (\ref{pAS}) to get $p_i^A$ in the stationary state, and $\nu_i^A=\frac{\sum_j a_{ij}p_j^A}{k_i^A}$, $\nu_i^B=\frac{\sum_j b_{ij}p_j^A}{k_i^B}$.
\section{Numerical simulations}
To validate our MMCA method, we carry out extensive numerical simulations and find a good agreement between the analytical and numerical results. In simulations, the uncorrelated configuration model (UCM) is used to generate each layer of the multiplex network, which follows a power law degree distribution~\cite{newman2001}. The number of nodes is set to be $N=10000$, the power exponent is $c=2.5$, the minimal and maximal degree are $k_{min}=3$ and $k_{max}=N^{-(c-1)}$ respectively. The nodes of two layers are randomly connected thus each node in one layer has a counterpart in the other layer. Initially, 10\% nodes are randomly selected as infected seeds and the remaining nodes are susceptible.

\begin{figure}
\begin{center}
\includegraphics[width=13.5cm]{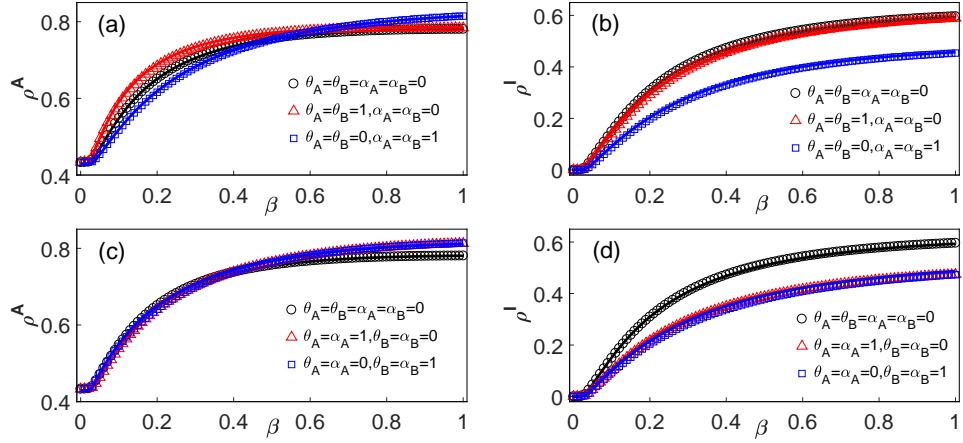}
\caption{The awareness density and the infection density obtained by the MMCA (lines) and Monte Carlo simulations (shapes) as a function of basic disease transmission rate $\beta$. (a) The awareness density under classical model(black circle), confirmation from infected neighbors (red triangle) and confirmation from aware neighbors (blue square). (b) The infection density under classical model, confirmation from infected neighbors and confirmation from aware neighbors. (c) The awareness density under classical model (black circle), confirmation from layer A (red triangle) and confirmation from layer B (blue square). (d) The infection density under classical model, confirmation from layer A and confirmation from layer B. Other dynamic parameters are set as $\lambda=0.24$, $\delta=0.6$, $\gamma=0.5$, $\mu=0.6$. The simulation results are obtained by averaging over 100 independent runs.}
\label{figure3}
\end{center}
\end{figure}

Fig.~\ref{figure3} demonstrates the density of aware nodes $\rho^A$ and infected nodes $\rho^I$ in the stationary state as a function of basic disease transmission rate $\beta$ under different parameters, where $\rho^I=\frac{1}{N}\sum_i p_i^{AI}$ and $\rho^A=\frac{1}{N}\sum_i (p_i^{AI}+p_i^{AS})$.
In Fig.~\ref{figure3} (a) and (b), three cases are compared, which are (1) information-disease interacting dynamics with no mutual confirmation, corresponding to $\theta_A=\theta_B=\alpha_A=\alpha_B=0$, (2) only the infection of neighbors are mutually confirmed to increase the information receiving rate, corresponding to $\theta_A=\theta_B=1$, $\alpha_A=\alpha_B=0$, and (3) only the awareness of neighbors are mutually confirmed to reduce the attenuation factor of disease transmission rate, corresponding to $\theta_A=\theta_B=0$, $\alpha_A=\alpha_B=1$. It can be seen that the when $\beta$ is smaller than $0.6$, the awareness density $\rho^A$ is the highest when there is mutual confirmation from infected neighbors. This is because the information transmission rate is directly enhanced due to the confirmation from infected neighbors. The awareness density is the lowest when there is mutual confirmation from the aware neighbors. This is because the disease transmission rate is reduced in this case, and the number of aware nodes informed by their infected counterparts decreases correspondingly. When $\beta$ is greater than $0.6$, the $\rho^A$ corresponding to $\alpha_A=\alpha_B=1$ is growing larger than the other cases. This is because when the disease transmission rate is large enough, the proportion of infected neighbors $\omega_i^A$ and $\omega_i^B$ become large and thus significantly enhance the information receiving rate of nodes, as defined in Eq. (1). As for the infection density $\rho^I$, it can be seen that for both cases with mutual confirmation, the infection is suppressed compared with the case with no mutual confirmation. The infection density is the lowest when the aware neighbors are mutually confirmed, corresponding to the case of $\theta_A=\theta_B=0$ and $\alpha_A=\alpha_B=1$. This is because the disease transmission rate is reduced directly. When the infected neighbors are mutually confirmed, the information receiving rate of nodes increases, resulting in the increase of aware nodes in layer A and decrease of infection in layer B. So $\rho^I$ for the case of $\theta_A=\theta_B=1$ and $\alpha_A=\alpha_B=0$ is smaller than that of the classical model.

In Fig.~\ref{figure3} (c) and (d), the effects of confirmation from layer A an layer B are demonstrated. It can be seen that neighbors of layer A (corresponding to $\theta_A=\alpha_A=1$ and $\theta_B=\alpha_B=0$) or layer B (corresponding to $\theta_A=\alpha_A=0$ and $\theta_B=\alpha_B=1$) have a similar impact on $\rho^A$, which is very close to that of the classical model when $\beta$ is small and a bit higher than that of the classical model when $\beta$ is greater than $0.6$. As for $\rho^I$, the effects of confirmation from layer A or layer B are very obvious, leading to a reduced infection density than the classical model. This is because the disease transmission rate is reduced and information transmission rate is enhanced at the same time, which together suppress the disease transmission. This result implies that if the neighbors in either layer can explicitly transmit their attitude and infection status, it will help to reduce the infection of epidemic disease. In the later part, we will discuss the impact of confirmation from layer A and layer B in detail.

\subsection{Effect of confirmation mechanism on the epidemic threshold}
\begin{figure}
\begin{center}
\includegraphics[width=13.5cm]{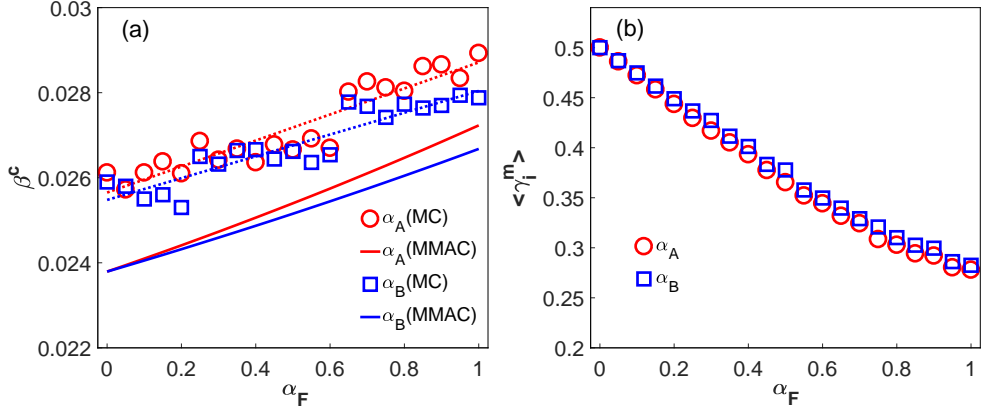}
\caption{The impact of information confirmation on the epidemic threshold. (a) The epidemic threshold as a function of confirmation strength $\alpha_F$. $F$ can be either $A$ or $B$. The result obtained by the MMCA (lines) and Monte Carlo (MC) simulations of susceptibility method (shapes) are demonstrated. (b) The average attenuation factor for disease transmission rate as a function of confirmation strength $\alpha_F$. Other dynamic parameters are set as: $\lambda=0.24$, $\delta=0.6$, $\gamma=0.5$, $\mu=0.6$.}
\label{figure4}
\end{center}
\end{figure}

According to the MMCA method, the epidemic threshold is strongly dependent on the matrix $H$, whose elements are $h_{ij}=[p_i^U+\gamma(1-\alpha_A\nu_i^A)(1-\alpha_B\nu_i^B)p_i^A]b_{ij}$. Here the $\nu_i^A$ and $\nu_i^B$ represent the proportion of aware neighbors in layer A and B respectively, and the parameter $\alpha_A$ and $\alpha_B$ represent the strength of confirmation from neighbors. If the information spreading does not outbreak when the information transmission rate $\lambda<\lambda_c$, then $\nu_i^A=\nu_i^B=0$ and the parameter $\alpha_A$ and $\alpha_B$ will have no impact on the epidemic threshold. Only when the information spreading outbreaks when $\lambda>\lambda_c$, the information spreading can suppress the disease spreading~\cite{wang2014}. In this case, $\nu_i^A\neq 0$ and $\nu_i^B\neq0$ and $\alpha_A$ and $\alpha_B$ work. On the other hand, as the infected individuals are very few near the epidemic threshold, $\omega_i^A\approx 0$ and $\omega_i^B\approx 0$. Then as described in Eq.(\ref{lamda}), the two parameters $\theta_A$ and $\theta_B$ representing the confirmation strength of $\omega_i^A$ and $\omega_i^B$ have no impact on $\lambda^{m}_i$ and the epidemic threshold. So we focus on the impact of $\alpha_A$ and $\alpha_B$ on the epidemic threshold $\beta_c$. Fig.~\ref{figure4} shows the epidemic threshold as a function of $\alpha$. The threshold obtained by our MMCA method and numerical simulations are compared. The simulated epidemic threshold is obtained by using a method called susceptibility~\cite{ferreira2012,shu2015}, where a $\chi$ is defined as
\begin{equation}\label{chi}
\chi=N\frac{\langle\rho^2\rangle-\langle\rho\rangle^2}{\langle\rho\rangle}.
\end{equation}
In the equation, $\rho$ is the infection density in the stationary state in one run, and $\langle\rho\rangle$ is the average over several runs. We implement 100 runs under each of the infection rate to get a $\chi$, and the infection rate that corresponds to the largest $\chi$ is identified as the simulated epidemic threshold. As shown in Fig.~\ref{figure4}, the epidemic threshold increases with $\alpha_F$, where $F$ is either A or B. The increment of $\beta_c$ is larger when $\alpha_A$ is applied than that of $\alpha_B$, which is due to the lower average attenuation factor for disease transmission $\langle\gamma_i^{m}\rangle$ of $\alpha_A$ as shown in Fig.~\ref{figure4} (b).

\begin{figure}
\begin{center}
\includegraphics[width=13.5cm]{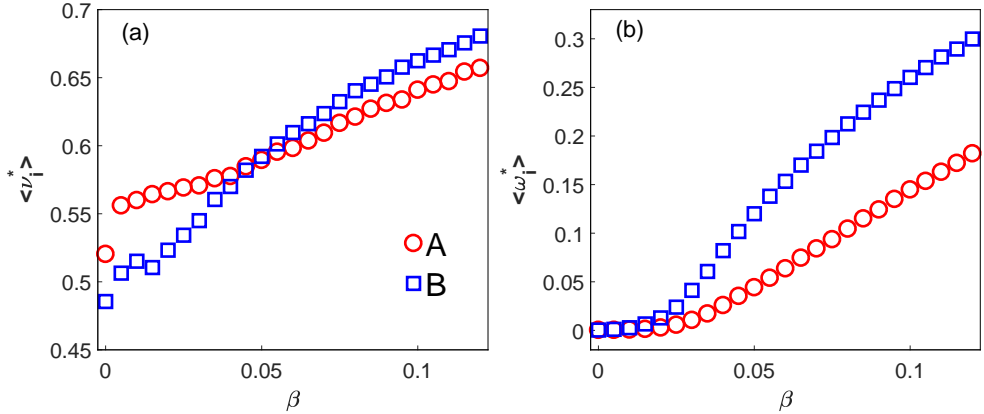}
\caption{The average proportion of aware or infected neighbors as as a function of basic disease transmission rate. (a) The average proportion of aware neighbors in layer A and B. (b) The average proportion of infected neighbors in layer A and B. In calculating $\langle\gamma_i^{m}\rangle$, we only take the AS nodes with at least one I neighbor into consideration, because only these nodes will be affected by $\gamma_i^{m}$. Similarly, we only take U nodes with at least one A neighbor into consideration when calculating $\langle\lambda_i^{m}\rangle$, because only these nodes will be affected by $\lambda_i^{m}$. The cross point at which $\langle v_i^B\rangle$ exceeds $\langle v_i^A\rangle$ is larger than $\beta_c$, which is the result of local effects in both information-spreading and disease-spreading. Other parameters are set as $\lambda=0.24$, $\delta=0.6$, $\gamma=0.5$, $\mu=0.6$, $\theta_A=\theta_B=\alpha_A=\alpha_B=0$. When any of the parameters $\theta_A$, $\theta_B$, $\alpha_A$, $\alpha_B$ is non-zero, $\lambda_i^m$ increases or $\gamma_i^m$ decreases, which do not change the local effect in spreading. }
\label{figure-new}
\end{center}
\end{figure}

The reason why the average $\gamma_i^{m}$ of $\alpha_A$ is lower than that of $\alpha_B$ can be explained by comparing the size of the average proportion of aware neighbors $\langle v_i^A\rangle$ in layer A and $\langle v_i^B\rangle$ in layer B . According to Eq.(\ref{gamma}), $\gamma_i^{m}=\gamma(1-\alpha_A\nu_i^A)(1-\alpha_B\nu_i^B)$. It can be seen from Fig.~\ref{figure-new} (a) that around the epidemic threshold $\beta_c$, $\langle v_i^A\rangle >\langle v_i^B\rangle$. Thus at a fixed value of $\alpha_A$ or $\alpha_B$, $\langle\gamma_i^{m}\rangle$ is smaller when $\alpha_A$ is applied than that of $\alpha_B$. When the disease transmission rate is under or around the epidemic threshold, the epidemic does not breakout and there are very few infected nodes. The aware nodes are mostly generated by the information spreading on layer A. As the spreading is from an aware node to its neighbors, the newly generated aware nodes are locally around the previous aware nodes. Thus $\langle v_i^A\rangle$ is relatively high for nodes with aware neighbors. While for nodes in layer B, as there are few infected nodes, their counterparts as aware nodes are correspondingly few. Meanwhile the aware nodes in layer A, if mapping to layer B, are randomly distributed in layer B because the two layers have no overlapping edges. Thus $\langle v_i^B\rangle$ is relatively small. When the disease transmission rate is above the epidemic threshold, the disease outbreaks and there are more and more infected nodes. In such stage, the proportion of aware neighbors for nodes in layer B exceeds that of nodes in layer A. This is because the nodes which have a possibility of being infected in layer B has at least one infected neighbors and thus one aware neighbors. For nodes in layer A, there is no such constraints on neighbors. So $\langle v_i^B\rangle$ is a bit greater than $\langle v_i^A\rangle$ when $\beta$ becomes larger. The cross point at which $\langle v_i^B\rangle$ exceeds $\langle v_i^A\rangle$ is larger than $\beta_c$. This means the local effect of information spreading dominates at first and then the coevolution of information-spreading and disease-spreading makes the relative proportion of aware neighbors in the two layers changes.

Similarly, we plot the average proportion of infected neighbors $\langle \omega_i^A \rangle$ in layer A and $\langle \omega_i^B\rangle$ in layer B  as a function of $\beta$ respectively. As can be seen in Fig.\ref{figure-new} (b) that $\langle \omega_i^B\rangle$ is greater than $\langle \omega_i^A\rangle$ under all disease transmission rate. This is because the infection spreads on layer B. As an infected node infects its neighbors, the newly infected nodes are locally around the infecting nodes. The infected nodes are locally clustered, resulting in a relatively large $\langle \omega_i^B \rangle$. Mapping these infected nodes to their counterparts in layer A, the infected nodes are scattered randomly. Thus $\langle \omega_i^A\rangle$ is relatively small.

These results imply that when the information on disease spreads only by relationships, it is local at the initial stage, which is not beneficial for suppressing the disease spreading. If at this stage the public health authorities can propagate the information on disease through mass media, it will help suppress the spreading of epidemics.

\begin{figure}
\begin{center}
\includegraphics[width=15cm]{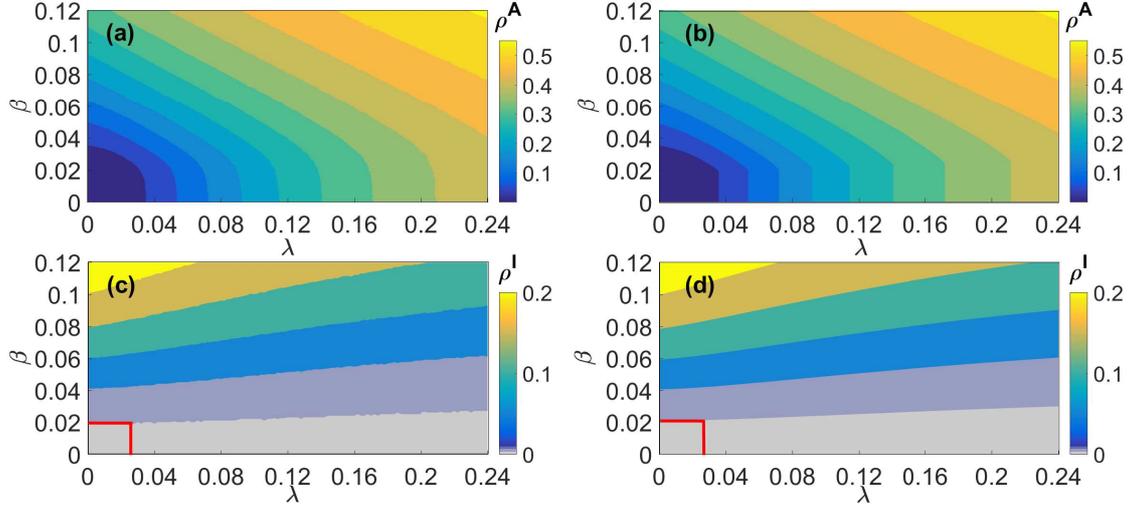}
\caption{The density of aware individuals $\rho^A$ and infected individuals $\rho^I$ obtained from Monte Carlo simulations and MMCA. (a) Simulated density of aware individuals. (b) Theoretical density of aware individuals. (c) Simulated density of infected individuals. (d) Theoretical density of infected individuals. In Monte Carlo simulations, each result is obtained by averaging over 100 runs. Other parameters are set as $\delta=0.6$, $\gamma=0.5$, $\mu=0.6$, $\theta_A=\theta_B=0$, $\alpha_A=\alpha_B=0.5$.}
\label{figure5}
\end{center}
\end{figure}

In Fig.~\ref{figure5} we compare the theoretical and numerical values of the density of aware individuals $\rho^A$ and infected individuals $\rho^I$ under different $\beta$ and $\lambda$.
It can be seen from Fig.~\ref{figure5} (a) and (b) that, with the increase of $\beta$ or $\lambda$, $\rho^A$ increases. In Fig.~\ref{figure5} (c) and (d), $\rho^I$ increases with $\beta$ and decreases with $\lambda$. When $\lambda<\lambda_c$, the information spreading does not break out thus has no effect on disease spreading. This corresponds to the plateau of $\beta_c$ marked in Fig.~\ref{figure5} (c) and (d). When $\lambda>\lambda_c$, the epidemic threshold $\beta_c$ increases with $\lambda$. This is because in $h_{ij}$, $\nu_i^A$ and $\nu_i^B$ represent the proportion of aware neighbors. Increasing the information transmission rate $\lambda$ can increase the value of $\nu_i^A$ and $\nu_i^B$, thus decrease the attenuation factor of disease transmission rate $\gamma_i^{m}$ and increases the epidemic threshold.
\subsection{Effect of different layers on suppressing the disease spreading}
\begin{figure}
\begin{center}
\includegraphics[width=15cm]{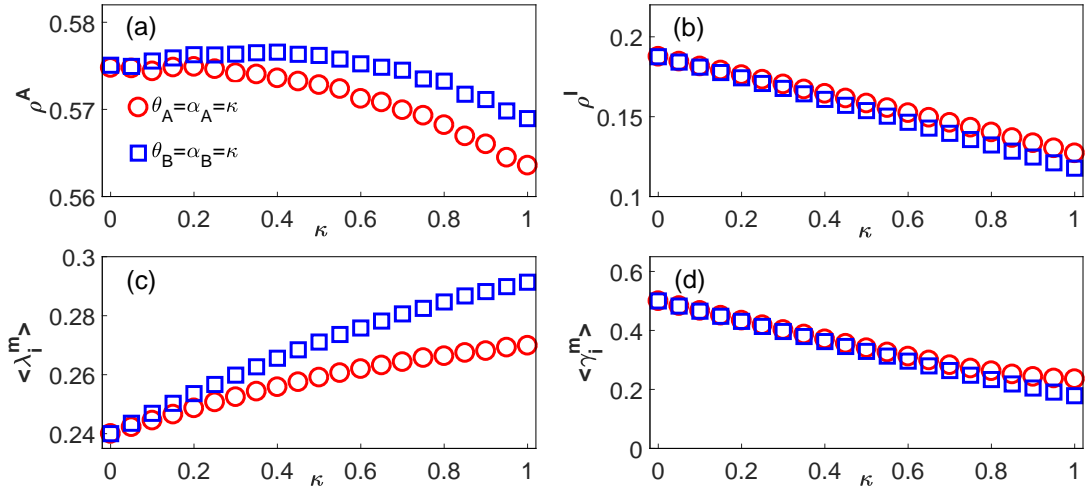}
\caption{Comparison of confirmation effect from layer A and layer B. (a) Effect of confirmation on $\rho^A$. (b) Effect of confirmation on $\rho^I$. (c) Effect of confirmation on average information transmission rate $\langle\lambda_i^{m}\rangle$. (d) Effect of confirmation on average attenuation factor $\langle\gamma_i^{m}\rangle$. In calculating $\langle\gamma_i^{m}\rangle$, we only take the AS nodes with at least one I neighbor into consideration, because only these nodes will be affected by $\gamma_i^{m}$. Similarly, we only take U nodes with at least one A neighbor into consideration when calculating $\langle\lambda_i^{m}\rangle$, because only these nodes will be affected by $\lambda_i^{m}$. Other parameters are set as $\delta=0.24$, $\delta=0.6$, $\beta=0.12$, $\mu=0.6$, $\gamma=0.5$.}
\label{figure6}
\end{center}
\end{figure}

In the proposed mutual confirmation mechanism, the infected and aware neighbors of communication layer and contact layer can affect the information receiving rate and disease transmission rate of nodes. Then the confirmation from which layer suppresses the disease spreading and promotes the information spreading more effectively is the question we are interested in. Fig.~\ref{figure6} demonstrates the $\rho^A$ and $\rho^I$ in the stationary state when the confirmation from layer A or layer B is applied.
It can be seen from Fig.\ref{figure6} (a) that when only the confirmation from neighbors in layer B is applied, the density of aware individuals is larger than that of layer A. This is because the average information transmission rate $\langle\lambda_i^{m}\rangle$ is higher when the confirmation from layer B is applied than that when only confirmation from layer A is applied, as can be seen in Fig.\ref{figure6} (c). The higher $\langle\lambda_i^{m}\rangle$ when the confirmation from layer B is applied is due to the higher $\langle\omega_i^B\rangle$ as demonstrated in Fig.\ref{figure-new} (b), which is a result of the locally clustered newly infected nodes in the disease-spreading process.

From Fig.~\ref{figure6} (b) we can see that the density of infected individuals is slightly lower when the confirmation from layer B is applied than that of layer A. This is because the average attenuation factor for disease transmission rate $\langle\gamma_i^{m}\rangle$ when the confirmation from layer B is applied is smaller than that of layer A, as can be seen from Fig.\ref{figure6} (d). So the lower $\rho^I$ of confirmation from layer B may due to the smaller $\gamma_i^{m}$ and the wider spreading of awareness $\rho^A$ as shown in (a).

The above results imply that the confirmation from contact layer can promote the diffusion of awareness more widely and suppress the disease more deeply than that of the communication layer. In real-world scenarios, for the public healthy authorities to control the diffusion of disease, let the individuals timely and accurately reveal their infected status and aware status, especially for those having real contacts in daily life, is effective to further suppress the epidemics.
\subsection{Impact of each confirmation information on suppressing the disease spreading}
\begin{figure}
\begin{center}
\includegraphics[width=15cm]{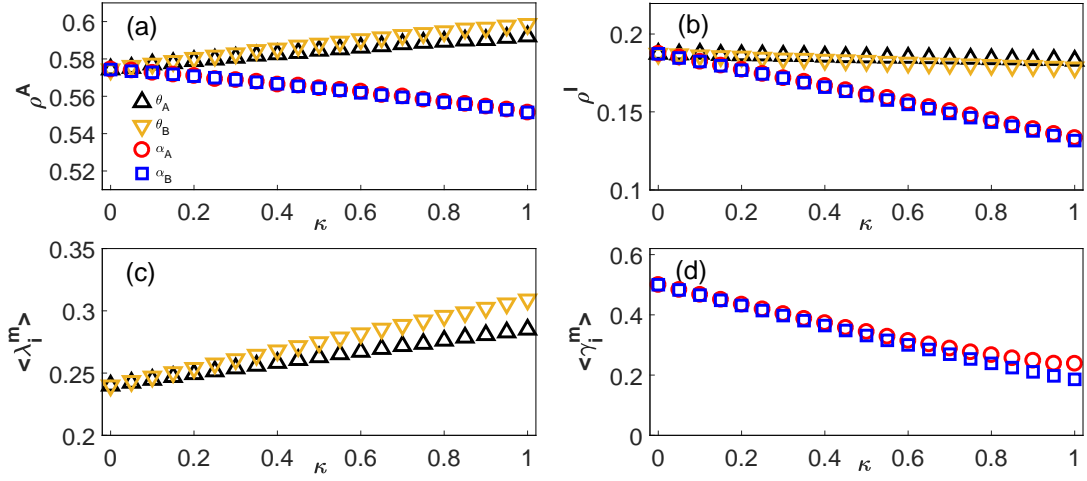}
\caption{Effect of the confirmation information. (a) Comparison of $\rho^A$ for each of the four confirmation information. (b) Comparison of $\rho^I$ for each of the four confirmation information.  When we study on one confirmation strength, other three confirmation strength parameters are set to be 0. E.g., in subgraph (a) the curve with blue triangle represents $\rho^A$ as a function of $\theta_A$, while other three confirmation strength parameters are set to be $\theta_B=\alpha_A=\alpha_B=0$. (c) Comparison of average information transmission rate $\lambda_i^{m}$ for $\theta$. (d) Comparison of average attenuation factor $\gamma_i^{m}$ for $\alpha$. In calculating $\langle\gamma_i^{m}\rangle$, we only take the AS nodes with at least one I neighbor into consideration, because only these nodes will be affected by $\gamma_i^{m}$. Similarly, we only take U nodes with at least one A neighbor into consideration when calculating $\langle\lambda_i^{m}\rangle$, because only these nodes will be affected by $\lambda_i^{m}$. Other parameters are set as $\delta=0.24$, $\delta=0.6$, $\beta=0.12$, $\mu=0.6$, $\gamma=0.5$.}
\label{figure7}
\end{center}
\end{figure}

In this part, we focus on the impact of each confirmation information, which are the $\nu_i^A$, $\nu_i^B$, $\omega_i^A$ and $\omega_i^B$, on promoting information diffusion and suppressing infection. To do this, we vary the confirmation strength of each confirmation information, which are $\theta_A$, $\theta_B$, $\alpha_A$ and $\alpha_B$ respectively, and investigate the density of aware individuals $\rho^A$ and infected individuals $\rho^I$. From Fig.\ref{figure7} (a) it can be seen that $\rho^A$ is the highest when $\theta_B$, the confirmation strength of the proportion of infected neighbors $\omega_i^B$ in layer B, is introduced. While $\theta_A$ is introduced, which is the confirmation strength of the proportion of infected neighbors $\omega_i^A$ in layer A , $\rho^A$ is lower than that of $\theta_B$. This is because the average information transmission rate $\langle\lambda^{m}_i\rangle$ is higher when $\theta_B$ is applied than that of $\theta_A$ as shown in Fig.~\ref{figure7} (c). The higher $\langle\lambda^{m}_i\rangle$ of $\theta_B$ is due to the larger $\omega_i^B$ shown in Fig.\ref{figure-new} (b), which is the result of local effect in disease spreading process. When $\alpha_A$ and $\alpha_B$ are applied respectively, $\rho^A$ is the lowest. This is because the non-zero $\alpha_A$ and $\alpha_B$ in the attenuation factor for disease transmission $\gamma_i^{m}$ reduce the disease transmission rate directly, thus leading to a decreased number of infected individuals as well as their counterparts as aware nodes.

In Fig.~\ref{figure7} (b) when $\alpha_A$ and $\alpha_B$ are applied respectively, $\rho^I$ is lower than that of $\theta_A$ and $\theta_B$. This is because the disease transmission rate is directly reduced when $\alpha_A$ or $\alpha_B$ is applied. The average attenuation factor for disease transmission $\langle\gamma_i^{m}\rangle$ of $\alpha_B$ is lower than that of $\alpha_A$ as shown in Fig.~\ref{figure7} (d), which leads to the lower $\rho^I$ of $\alpha_B$. The smaller $\langle\gamma_i^{m}\rangle$ when $\alpha_B$ is introduced is the result of larger proportion of aware neighbors $\langle\nu_i^B\rangle$  in layer B as shown in Fig.\ref{figure-new} (a) when the basic disease transmission rate is far above the threshold, which is the result of the local effect in disease spreading process. The reason $\rho^I$ of $\theta_A$ is higher than that of $\theta_B$ is that $\theta_A$ results in a lower $\rho^A$ than $\theta_B$, thus the suppressing effect is smaller than that of $\theta_B$.

The above results indicate that in suppressing disease spreading, the rank for the effect of confirmation from neighbors is $\alpha_B>\alpha_A>\theta_B>\theta_A$. This implies that directly reducing the disease transmission rate, other than promoting awareness spreading, can suppress the disease spreading more effectively. In addition, the local effect in disease spreading makes the average proportion of infected neighbors in the physical contact layer larger than that of communication layer, which leads to the impact of $\theta_B>\theta_A$.
\section{Conclusion and Discussion}
In the early stage of epidemics, people are not certain about the credibility of information, thus may seek for confirmation from multiple sources. The infection of neighbors from both the communication layer and the contact layer will convince people the prevalence of disease, thus making the transmission of information on epidemics more easily. Meanwhile if there is a large proportion of acquaintances that have accepted the information, people are more likely to adopt protective measures to reduce the risk of infection. Based on this assumption, we articulate the information-disease interacting spreading model with inter-layer mutual confirmation mechanism on multiplex networks. By using the microscopic Markov chain method, we analytically predict the epidemic threshold and infection density in stationary state, which agree well with the simulation results.

We find that the confirmation from aware neighbors in layer A can enhance the epidemic threshold more than the confirmation from aware neighbors in layer B. This is because around the epidemic threshold there are few infected nodes and the spreading of information dominates the diffusion of awareness. The local effect of information spreading makes the proportion of aware neighbors in layer A larger than that of layer B. On the other hand, the infected neighbors of either layer A or B has no impact on the epidemic threshold, because around the epidemic threshold the number of infected neighbors is approaching zero.

As for the effect of suppressing disease, when independently apply one of the four confirmation information, the reduction of $\rho^I$ can be ordered as $\Delta_{\alpha_B}>\Delta_{\alpha_A}>\Delta_{\theta_B}>\Delta_{\theta_A}$. There are two reasons for this ranking. Firstly, reducing the disease transmission rate directly suppresses the epidemic more than enhancing the awareness spreading does. Secondly, the local effect in disease spreading makes the proportion of infected neighbors in layer B, as well as the proportion of aware neighbors of layer B, larger than that of layer A, thus leading to the larger inhibitory effect of confirmation from layer B. From the perspective of layers, confirmation from the aware and infected neighbors in contact layer can result in a lower infection density in the stationary state than that of the confirmation from the communication layer, which is also due to the local effect in disease-spreading and information-spreading.

The results in this work imply that when epidemic outbreaks, encouraging people to explicitly express their infected status and aware attitude is helpful to reduce the infection in the whole population. While the spreading of disease and information has a local effect, the heath authority's announcement of the information on epidemics to the public is important to suppress the disease. When the anti-epidemic resource is limited, investing the limited resource into real-world activities, such as distributing protective goods or medicines, cutting off disease transmission path, is more effective than propagating online.

\section*{Acknowledgement}
This work is supported by the National Natural Science Foundation of China (No. 61802321, 11975099), the Sichuan Science and Technology Program (No. 2020YJ0125), the Natural Science Foundation of Shanghai (No. 18ZR1412200) and the Science and Technology Commission of Shanghai Municipality (Grant No. 14DZ2260800).

\end{document}